\documentclass{article}
\usepackage{spconf,amsmath,graphicx}

\usepackage{numprint}
\usepackage{cleveref}
\crefformat{equation}{(#2#1#3)}
\crefformat{figure}{Fig.~#1}

\usepackage{booktabs}
\usepackage{balance}

\def\MOS{{$\mathrm{D}_\mathrm{MOS}$}}
\def\MOSSIGBAK{{$\mathrm{D}_\mathrm{OVR,SIG,BAK}$}}
\def\APE{{$\mathrm{D}_\mathrm{T60,C50}$}}

\title{Speech MOS multi-task learning and rater bias correction}
\twoauthors
  {Haleh Akrami\sthanks{This work was done while Haleh Akrami was an intern at Microsoft Research Labs in Redmond, WA, USA.}}
	{University of Southern California\\
	Department of Biomedical Engineering\\
	Los Angeles, CA, USA}
  {Hannes Gamper}
	{Audio and Acoustics Research Group\\
	Microsoft Research\\
	Redmond, WA, USA}
\begin{document}
\ninept

\maketitle
\begin{abstract}
Perceptual speech quality is an important performance metric for teleconferencing applications. The mean opinion score (MOS) is standardized for the perceptual evaluation of speech quality and is obtained by asking listeners to rate the quality of a speech sample. Recently, there has been increasing research interest in developing models for estimating MOS blindly. Here we propose a multi-task framework to include additional labels and data in training to improve the performance of a blind MOS estimation model. Experimental results indicate that the proposed model can be trained to jointly estimate MOS, reverberation time (T60), and clarity (C50) by combining two disjoint data sets in training, one containing only MOS labels and the other containing only T60 and C50 labels. Furthermore, we use a semi-supervised framework to combine two MOS data sets in training, one containing only MOS labels (per ITU-T Recommendation P.808), and the other containing separate scores for speech signal, background noise, and overall quality (per ITU-T Recommendation P.835). Finally, we present preliminary results for addressing individual rater bias in the MOS labels.
\end{abstract}
\section{Introduction}
\label{sec:intro}

Speech quality is a key performance metric of modern telecommunication systems. Subjective speech quality can be measured by asking human judges to rate the perceived quality of a speech sample on a 5-point Likert scale ranging from 1 (``bad'') to 5 (``excellent'') and then averaging several ratings to derive a Mean Opinion Score (MOS)~\cite{P808}. While crowd-sourcing offers a way to obtain MOS ratings at scale~\cite{P808,naderi2020open}, it remains a costly and time-consuming approach to measure speech quality. Recently, there has been substantial research interest in estimating speech quality automatically and blindly using computational models~\cite{gamper2019intrusive,reddy2021dnsmos,mittag2021nisqa,faridee2022predicting}, including a special challenge at Interspeech 2022 on Non-intrusive Objective Speech Quality Assessment (NISQA)~\cite{NISQA2022}. While computational models can be a cost-effective way to obtain speech quality estimates, they are typically data-driven and thus require large amounts of labeled training data, usually obtained via crowd-sourcing~\cite{naderi2020open,mittag2021nisqa}.
Here we investigate whether MOS training data can be augmented using complementary labels and data, and whether the quality of crowd-sourced MOS ratings can be improved by estimating and correcting rater bias. Our approach builds on prior work on multi-task learning with partially labeled data as well as research into addressing quality issues in crowd-sourced data. 
We assume that related labels e.g., acoustic parameters affecting speech perception (reverberation time and clarity) or additional MOS labels rating signal and background quality, may aid MOS estimation.

\subsection{Multi-task learning with missing labels}
Through sharing data and computational resources, multi-task learning (MTL) tries to develop a single model for numerous tasks with higher efficiency and generalization. Most MTL techniques rely on expensive fully labeled datasets. However, in practice, data sets may exhibit missing labels, or there may be opportunities to increase training data by combining multiple data sets that share a subset of labels. 
Durand et al.\ \cite{durand2019learning} combined neural graph networks (GNN) with curriculum learning-based strategies to predict missing labels for multi-task classification with missing labels. They used GNN to capture label dependency, used label uncertainty to trim the data in each epoch, and only trained on reliable labels. Kundo et al.\ \cite{kundu2020exploiting} used the idea of temperature to not consider the unannotated negative labels as hard negatives but as soft labels (positive/negative). Each unannotated label is assigned a unique temperature value based on its confidence in its hard label. Recently Baruch et al.\ \cite{ben2022multi} targeted the partial label problem by using focal loss and assumed all unlabeled data samples have negative labels based on the prior information. They used a focal loss for handling class imbalance and focusing on hard samples. From the unlabeled samples, they only used the samples with the highest likelihood in each iteration. Similarly, Kim et al.\ \cite{kim2022large} used a trimming strategy after assigning negative labels to unlabeled samples.

In the context of multi-task learning with missing labels for a regression problem,  Deng et al.\ \cite{deng2020multitask} used a teacher to predict a soft label for incomplete label entries using a temperature-scaling loss. They use the soft labels and the ground truth to train the student model with a distillation loss. Knowledge distillation for regression is not as common as for classification. To use knowledge distillation, they transform the regression task into a classification task by discretizing the continuous values. Then, they use the temperature to control the smoothness of the soft labels.

\subsection{MOS rater bias correction}
The question of how rater effects, including rater bias, affect performance ratings is a well-established area of research~\cite{wolfe2004identifying}. With crowdsourcing becoming a ubiquitous paradigm for retrieving data from humans to improve machine intelligence, controlling the quality of crowd work and mitigating the effect of workers' bias continues to be a challenge. 
For classification problems, crowdsourcing datasets often rely on majority voting, which treats all annotators as equally reliable and grants them equal votes. However, this approach ignores information contained in noisy labels, including the certainty of the integrated label. To consider the annotators' skills and variable data difficulty, Dawid and Skene parameterized annotators' reliability using their error rates and modeled the annotations as noisy observations of the latent ground truth~\cite{dawid1979maximum}.
 
Several extensions to this DS method were developed for a more reliable estimation of the bias~\cite{whitehill2009whose,zhou2012learning,li2019exploiting}. Liu et al.\ \cite{liu2012variational} transformed the problem of crowdsourcing into a standard inference problem using a variational inference perspective. Bayesian extensions of DS have also been developed to mitigate the issue of having too few labels per worker to estimate their quality reliably. They assume each worker belongs to a particular community, where the worker's confusion matrix is similar to (a perturbation of) the community's confusion matrix~\cite{venanzi2014community}. Recently, learning from crowdsourcing extended to deep neural networks. Albarqouni et al.\ estimate each annotator's sensitivity and specificity using the expectation maximization (EM) algorithm for a neural network~\cite{albarqouni2016aggnet}. An alternative approach is to use weighted majority voting by using the certainty of the integrated label as the weight of each instance~\cite{jiang2021learning,sheng2017majority}. Fornaciari et al.\ defined soft labels as probability distributions over the labels given by the annotators rather than using one-hot encodings with a single correct label and measured the divergence between probability distributions as the loss function~\cite{fornaciari2021beyond}.
Most of these methods are related to classification problems and use a cross-entropy loss. However, speech MOS estimation is typically treated as a regression problem~\cite{reddy2021dnsmos,faridee2022predicting}. 
In a listening test, participants' subjective responses to samples, e.g., ratings of the perceived speech quality, may be affected by individual differences in the internal scale or reference participants use to measure or rate their perceptual experience. Linear mixed-effects models can be used to account for these individual differences in the statistical analysis of subjective data with multiple responses per participant~\cite{Magezi2015}. Rosenberg and Ramabhadran suggest using normalized-rank normalization to correct rater bias when analysing or comparing listening test results~\cite{rosenberg2017bias}. Raymond and Viswesvaran propose least-squares methods for correcting the effect of individual raters on average performance scores~\cite{raymond1993least}. However, their methods involve matrix formulations that become impractical for large crowd-sourced datasets with thousands of raters. ITU-T Recommendation P.1401 proposes methods to address differences between MOS labels obtained in separate experiments, e.g., to compare listening test results from two separate laboratories~\cite{P1401}.

\begin{figure}
\caption{Neural network architecture.}
\label{fig:architecture}
  \centering
  \centerline{\includegraphics[width=8.5cm]{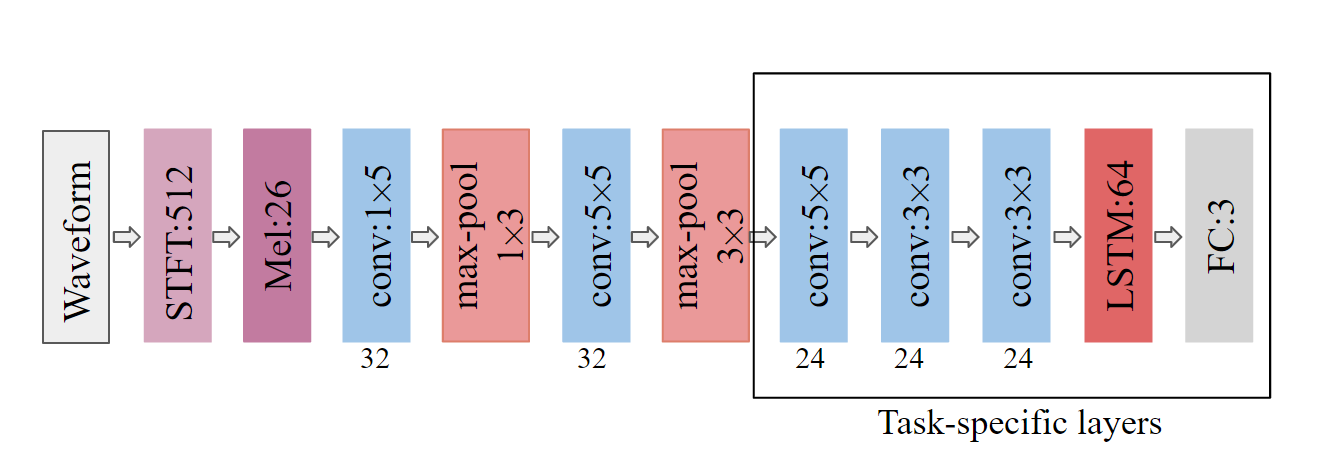}}
\end{figure}

\subsection{Contributions}
We aim to learn speech MOS in a multi-task learning (MTL) framework to leverage useful information in related tasks, increase the available training data size by combining heterogeneous data sets, and improve model generalization. 
Our contribution in this paper is two-fold: 
 (i) we propose a semi-supervised multi-task framework for combining disjoint data sets or data sets with partially missing labels to estimate multiple labels jointly, and (ii) we show preliminary results for two empirical methods to estimate and correct rater bias in speech MOS labels.

\section{Data}
\label{sec:data}

\subsection{MOS data}
The MOS training set consists of speech samples labeled by human raters using crowd-sourcing. The set contains a total of \numprint{418197} speech samples, with most processed by noise suppression models, including contestants in the Interspeech 2020 Deep Noise Suppression Challenge~\cite{reddy2020interspeech}. \numprint{274466} of the training samples are rated following ITU-T Recommendation P.808 and contain a single MOS label per file~\cite{naderi2020open}. This dataset is referred to as \MOS. 
\numprint{143731} of the training samples are rated following ITU-T Recommendation P.835 and contain three MOS labels per file, for overall (OVR), signal (SIG), and background noise quality (BAK), respectively~\cite{naderi2020subjective}. This dataset is referred to as \MOSSIGBAK.

As the test set, we use a portion of the blind set of the ICASSP 2022 deep noise suppression challenge~\cite{dubey2022icassp}. The test set contains \numprint{48104} samples, obtained from 1718 unique files processed with various noise suppression models including contestants in the ICASSP 2022 challenge and labeled by at least five human raters each.

\subsection{Acoustic parameter data}
The training set for acoustic parameter estimation contains a total of \numprint{200000} samples.
It consists of clean speech samples taken from the ``train'' portion of the TIMIT set~\cite{garofolo1993darpa} convolved with room impulse responses (RIRs) randomly selected from a variety of public and proprietary data sets. In addition, three types of noise are simulated, i.e., white Gaussian noise, ambient noise, and babble noise. Every resulting reverberant, noisy speech sample is labeled with the reverberation time (T60) and clarity (C50) derived from the RIR used~\cite{gamper2020blind}. 
The test set is created using the ``test'' portion of TIMIT and convolved with RIRs not contained in the training set, with additional Gaussian and babble noise, resulting in a total of \numprint{49920} samples. A detailed description of the acoustic parameter data sets and labels is provided in \cite{gamper2020blind}. The dataset is referred to here as \APE.

\section{Proposed method}

\begin{table*}
\caption{Model performance for MOS, T60 and C50 estimation.}\label{RCC_AP}
\centering
\begin{tabular}{|l|l|c|c|c|c|c|c|c|}
\hline
\bfseries Model & \bfseries \# parameters & \bfseries Training data & \multicolumn{3}{c|}{\textbf{Pearson correlation}} & \multicolumn{3}{c|}{\textbf{Root-mean-squared error}} \\
\hline
 &  & & \bfseries MOS & \bfseries C50 &\bfseries T60 & \bfseries MOS & \bfseries C50 &\bfseries T60\\
\hline
\emph{single} & 140k per task& \MOS{}, \MOSSIGBAK{}, \APE{} & \textbf {76.57\%} & 96.72\% & 89.98\%  & \textbf {0.5853} & 3.0101 & 0.5680\\ \hline
\emph{multi} & 140k & \MOS{}, \MOSSIGBAK{}, \APE{} & 74.82\% & 96.53\% & 89.23\%  & 0.6046 & 3.1785 & 0.5819\\ \hline
\emph{multi,split} & 258k & \MOS{}, \MOSSIGBAK{}, \APE{} & 75.92\% & 96.65\% & 90.09\%  & 0.5923 & 3.0067 & 0.5617\\ \hline
\emph{multi,split,W}& 258k & \MOS{}, \MOSSIGBAK{}, \APE{} & 76.42\% & 96.79\% & \textbf {90.77}\%  & 0.5876& 3.131 &\textbf { 0.5475} \\ \hline
\emph{multi,split,semi}& 258k & \MOS{}, \MOSSIGBAK{}, \APE{} & 75.96\% & \textbf {96.93}\% & 89.35\%  & 0.5908 & \textbf {2.9815}&  0.5778\\ \hline
\end{tabular}
\end{table*}

\subsection{Semi-supervised multi-task framework}
 

The proposed framework is built on a single-task model consisting of five convolutional layers followed by an LSTM layer~\cite{faridee2022predicting}. The model architecture is shown in \cref{fig:architecture}. The linear output layer produces one estimate per task for each sample. 

The model can be trained in a single-task manner on a single dataset containing a single label per sample, e.g., MOS. We refer to this model as \emph{single}.
The purpose of the proposed multi-task framework is to combine multiple heterogeneous datasets in training, specifically a dataset containing a single MOS label per sample (\MOS), a dataset containing a separate overall, signal, and background quality MOS label per sample (\MOSSIGBAK), and a dataset containing a reverberation time (T60) and clarity (C50) label per sample (\APE). 

The baseline architecture can be trained to estimate multiple labels per sample by using an output layer with one output node per task and a shared cost function, e.g., mean-squared error averaged over the output labels. The training data can be a set with complete labels (i.e., all samples have labels for all tasks, e.g., \MOSSIGBAK), or partial labels (i.e., some samples are missing some labels, e.g., if \MOS{} is combined with \MOSSIGBAK{} or \APE). Missing labels can be handled by simply ignoring their contribution in the loss function. We refer to this model as \emph{multi}.

To improve the ability of the network to learn task-specific features, the later layers of the network can be split into task-specific branches and trained separately for each task (cf.\ \cref{fig:architecture}). During training, for samples with missing labels only the task-specific layers for which labels are available are updated. This architecture is referred to as \emph{multi,split}.

As proposed by Deng et al.~\cite{deng2020multitask}, missing labels in a multi-task regression problem can be replaced with estimated or pseudo-labels in a semi-supervised framework. Here we obtain pseudo-labels for each task using a teacher model trained for that task. The teacher model can be trained separately for each task (e.g., using \emph{single}) or using a multi-task framework (e.g., using \emph{multi} or \emph{multi,split}). The student network is then trained in a multi-task fashion on the ground-truth and pseudo-labels. To improve performance for harder tasks, a weighted loss can be used, giving higher weight to more difficult tasks (see \cref{RCC_AP}, \emph{multi, split, W}).
To address the issue of noise inherent in pseudo-labels, a trimming strategy is employed~\cite{kim2022large},
whereby a portion of data with the highest loss is removed in each epoch. The trimming percentage can be tuned using a validation set. 

\subsection{Rater effect estimation and correction}
Similarly to prior work~\cite{raymond1993least,rosenberg2017bias}, we hypothesize that speech MOS rater bias can be estimated and corrected by comparing the ratings of an individual rater with the average ratings (or MOSs) of the samples rated by that rater. 
We propose two methods for addressing speech MOS rater bias; a simple bias removal, and a rating correction using a least squares linear fit.

For the bias removal, we assume that the average rating of an individual rater exhibits a bias relative to the true MOS of the rated samples. Related concepts are the rater \emph{leniency} or \emph{harshness}~\cite{raymond1993least,wolfe2004identifying}. For a single speech sample $s$, the MOS $M_s$ is defined as 
\begin{equation}
\label{eq:MOS}
    M_s = \frac{1}{\mathrm{N}_s}\sum_{i=1}^{\mathrm{N}_s}{r_{s,i}},
\end{equation}
where $r_i$ are the ratings of $\mathrm{N}_s$ raters. Assuming $\mathrm{N}_s\gg1$, we can obtain a MOS estimate $\hat{M}_{s,j}$ after removing rater $j$ as
\begin{equation}
\label{eq:MOSj}
    \hat{M}_{s,j} = \frac{1}{\mathrm{N}_s-1}\sum_{i=1;i\ne j}^{\mathrm{N}_s}{r_{s,i}}.
\end{equation}
The bias $b_j$ of rater $j$ can be estimated as
\begin{equation}
\label{eq:biasj}
    b_j = \frac{1}{\mathrm{S}_j} \sum_{s=1}^{\mathrm{S}_j}{\left(r_{s,j} - \hat{M}_{s,j}\right)},
\end{equation}
where $\mathrm{S}_j$ is the number of speech samples rated by rater $j$.
The \emph{unbiased} MOS estimate $\tilde{M}_s$ is obtained by replacing all ratings $r_s$ in \cref{eq:MOS} with the bias-corrected ratings $\tilde{r}_{s,j} = r_{s,j} - b_j$. 
Note that we did not repeat the bias estimation steps in \cref{eq:MOSj} and \cref{eq:biasj} with updated, bias-corrected ratings. Developing an iterative bias estimation procedure is left for future work.

Besides exhibiting bias, raters may have a tendency to either restrict ratings to a limited portion of the rating scale or assign extreme values. Related concepts are rater \emph{centrality} and \emph{extremism}~\cite{wolfe2004identifying}. We propose approximating these effects using a linear model
\begin{equation}
\tilde{r}_j = a_j r_j + b_j,
\end{equation}
solved in the least-squares sense as
\begin{equation}
    \min_{a_j,b_j} \sum_{s=1}^{\mathrm{S}_j}{\left( (a_j r_j + b_j) - \hat{M}_{s,j}\right)^2}.
    \label{eq:linfit}
\end{equation}

\section{Experimental results}
Experiments were performed for the proposed multi-task framework on combinations of the data sets described in \cref{sec:data}.
Model training was performed using a mean-squared error loss with an Adam optimizer, a batch size of 256, and a learning rate scheduler with an initial rate of 0.001. We report (i) Pearson’s correlation coefficient (PCC) and (ii) the root-mean-squared error (RMSE) between estimated and ground-truth MOS as performance metrics.

\subsection{Multi-task MOS, T60, and C50 estimation}
We combine data sets \MOS{}, \MOSSIGBAK{}, and \APE{} to train models using the proposed multi-task framework. 
\Cref{RCC_AP} shows results for various models. For MOS estimation, the single-task model outperforms all multi-task models. However, C50 and T60 see a small benefit from multi-task training on MOS data that does not contain any T60 or C50 labels. This indicates that the proposed model can be used to jointly estimate all three tasks with fewer parameters than three separate single-task models.

\begin{table*}
\caption{Model performance for overall (OVR), signal (SIG), and background (BAK) quality MOS estimation. }
\label{RCC_SIGBAK}
\centering
\begin{tabular}{|l|l|c|c|c|c|c|c|c|}
\hline
\bfseries Model & \bfseries \# parameters & \bfseries Training data & \multicolumn{3}{c|}{\textbf{Pearson correlation}} & \multicolumn{3}{c|}{\textbf{Root-mean-squared error}} \\
\hline
 &  & & \bfseries OVR & \bfseries SIG &\bfseries BAK & \bfseries OVR & \bfseries SIG &\bfseries BAK\\
\hline

\emph{single} & 140k per task & \MOSSIGBAK{} & 75.56\% & 73.66\% & 70.05\% & 0.6111 & 0.5788 & 0.6045 \\ \hline
\emph{multi} & 140k & \MOSSIGBAK{} & 76.13\% & 75.33\% & 70.75\% & 0.6090 & 0.5608 &  0.5932 \\ \hline
\emph{multi} & 140k & \MOS{}, \MOSSIGBAK{} & 76.47\% & 75.08\% &70.49 \%  & 0.5995 & 0.5644 &0.5905 \\ \hline
\emph{multi,split} & 258k & \MOSSIGBAK{} & 76.74\% & 76.11\% & 71.46\%   & 0.5925 &  0.5570 & 0.5922\\ \hline
\emph{multi,split}  & 258k & \MOS{}, \MOSSIGBAK{}  & \textbf {77.31\%} & 74.78\% & 71.09\%   & \textbf {0.5813} & 0.5665 & 0.5944\\ \hline
\emph{multi,split,semi}  & 258k & \MOS{}, \MOSSIGBAK{} & 77.06\% &\textbf {76.13\%} & \textbf {71.89\%} &0.5846 &\textbf {0.5524} & \textbf {0.5886} \\ \hline

\end{tabular}
\end{table*}

\subsection{Multitask OVR, SIG, and BAK MOS estimation}
Combining data sets \MOS{} and \MOSSIGBAK{} allows leveraging information from presumably very closely related tasks, i.e., overall, signal, and background noise quality ratings. The results in \cref{RCC_SIGBAK} indicate that the multi-task framework benefits all three tasks, with the multi-task models outperforming the single-task baselines. Similarly to multi-task T60 and C50 estimation, the multi-task framework seems to successfully exploit information in the \MOS{} data set for SIG and BAK estimation, even though \MOS{} does not contain any SIG or BAK labels. 


\subsection{Rater bias estimation and correction}
\begin{figure}
  \centering
  \includegraphics{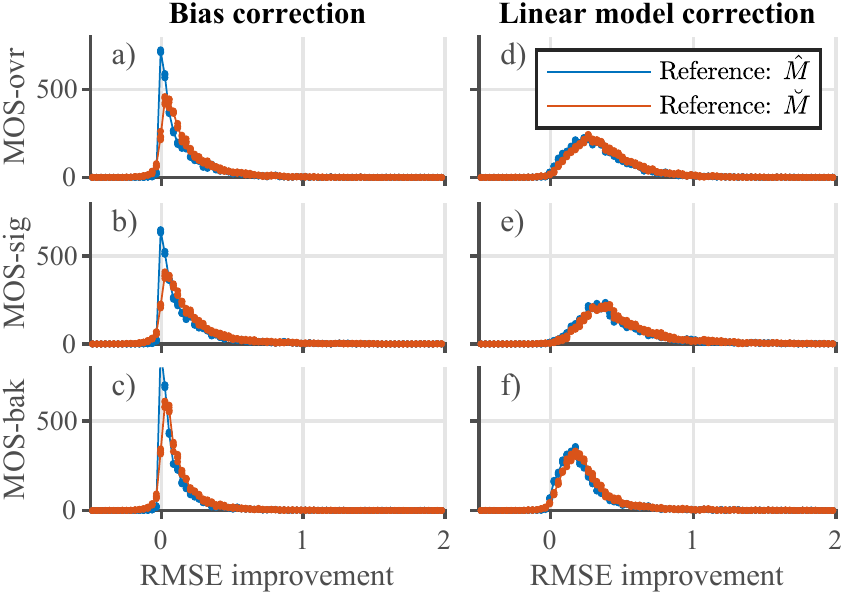}
\caption{Histogram of root-mean-squared error (RMSE) improvements per rater compared to MOS estimates $\hat{M}_{s,j}$ (blue) and $\breve{M}_s$ (red) for OVR (a,d), SIG (b,e), and BAK MOS (c,f), using bias correction (a-c) and linear model correction (d-f). Dots indicate the results of individual cross-validation runs, lines show the mean.}
\label{fig:bias-correction}
\end{figure}

To test the proposed rater effect estimation and correction, we carried out experiments on MOS labels of \numprint{29294} speech samples, obtained from at least 10 raters each, i.e., $\mathrm{N}_s \ge 10$. The set contains a total of 3363 unique raters that each have rated at least five speech samples. For each sample, $\hat{M}_{s,j}$ is calculated for rater $j$ via \cref{eq:MOSj} using four randomly selected raters. The remaining $\mathrm{N}_s - 5$ raters are used to estimate a second \emph{hold-out} MOS estimate $\breve{M}_s$ from a minimum of five raters excluding the rater $j$ and the four random raters used to calculate $\hat{M}_{s,j}$. This process is repeated five times for cross-validation. \cref{fig:bias-correction} shows the distribution of improvements of the unbiased ratings $\tilde{r}_j$ over the raw ratings $r_j$ with respect to MOS estimates $\hat{M}_{s,j}$ (blue) and $\breve{M}_s$ (red), for the proposed bias correction methods and overall, signal, and background quality MOS labels (OVR, SIG, BAK). As can be seen, both methods seem to improve rater performance for most raters in terms of the root-mean-squared error $\mathrm{RMSE_j} = \sqrt{\mathrm{S}_j^{-1}\sum^{\mathrm{S}_j}{\left( r_{s,j} - \breve{M}_s \right)^2}}$.

To gauge the potential effect of bias correction on MOS estimation performance, we evaluated using unbiased MOS estimates separately in training  and testing. However, for the training set, rater IDs were only known for 8190 samples, so the effect on model performance was expected to be minimal. For the test set, all samples had associated rater IDs. Bias correction was only applied to raters with a minimum of five rated samples, i.e., to 392 raters in the MOS training set and 2279 raters in the MOS test set. \cref{tab:debias_training} shows the results of correcting bias for 8190 samples in the training set. As can be seen, there is a small improvement in performance. A larger improvement may be possible by correcting bias for a larger portion or all training samples if rater IDs are available. \Cref{tab:debias_testing} shows that correcting for rater effects in the test samples yielded lower model estimation errors. However, more labels per sample would be needed to determine whether this is a result of the bias correction reducing label noise. Therefore, these results should be interpreted as a preliminary proof of concept.

\begin{table}
\caption{Training data rater effect correction for model \emph{single} trained on data set \MOSSIGBAK{}.}
\label{tab:debias_training}
\centering
\begin{tabular}{|c|c|c|}
\hline
\bfseries \bfseries Method & \bfseries MOS PCC & \bfseries MOS RMSE\\\hline

none & 75.56 \% &  0.6111\\ \hline 
bias correction & \textbf{75.93}\% & 0.6081\\ \hline
linear model correction & 75.35\% & \textbf{0.6030} \\ \hline

\end{tabular}
\end{table}

\begin{table}
\caption{Test data rater effect correction for model \emph{multi,split,semi} trained on data sets \MOS{} and \MOSSIGBAK{}.}
\label{tab:debias_testing}
\centering
\begin{tabular}{|c|c|c|c|}
\hline
\bfseries Method & \bfseries MOS PCC & \bfseries MOS RMSE \\\hline

none & 77.02\% & 0.5861 \\ \hline
bias correction & \textbf{79.98}\% & 0.5345\\ \hline
linear model correction & 79.77\% & \textbf{0.3921}\\ \hline

\end{tabular}
\end{table}

\section{Conclusions}

We proposed a multi-task framework with task-specific layers and a semi-supervised student--teacher network for MOS estimation. The experimental results show that the model can be trained with heterogeneous data sets with missing labels. In most cases, the proposed multi-task framework improves Pearson correlation and root-mean-squared error compared to single-task models trained for the individual tasks while using fewer trainable parameters. This suggests that the proposed framework successfully exploits information from heterogeneous training data.
Finally, we present preliminary results for addressing rater effects in  MOS labels. The proposed bias and linear model corrections show promise for improving the quality of crowd-sourced MOS labels as well as MOS estimation performance.

\vfill\pagebreak

\bibliographystyle{IEEEbib_abbr}
\bibliography{IEEEbib}

\end{document}